\documentclass[pra, onecolumn]{revtex4}
\usepackage{graphicx}

\usepackage{ulem}
\usepackage[section]{placeins}
\usepackage{amsmath}
\usepackage{amssymb}
\usepackage{dsfont}
\usepackage{natbib}

\usepackage{subfiles}
\usepackage{graphicx}
\usepackage{array}
\usepackage{calrsfs}
\usepackage{hhline}
\usepackage{relsize}

\newcolumntype{C}[1]{>{\centering\let\newline\\\arraybackslash\hspace{0pt}}m{#1}}
\begin{document}

\title{Robust super-resolution depth imaging via a multi-feature fusion deep network}

\author{Alice Ruget$^{1}$, Stephen McLaughlin$^{1}$, Robert K.~Henderson$^{2}$, Istvan Gyongy$^2$, Abderrahim~Halimi$^{1}$, Jonathan Leach$^{1,*}$}
\address{$^1$School of Engineering and Physical Sciences, Heriot-Watt University, Edinburgh, EH14 4AS, UK}
\address{$^2$School of Engineering, Institute for Integrated Micro and Nano Systems, The University of Edinburgh, Edinburgh, EH9 3FF, UK}
\address{$^*$j.leach@hw.ac.uk}

\begin{abstract}
The number of applications that use depth imaging is increasing rapidly, e.g. self-driving autonomous vehicles and auto-focus assist on smartphone cameras.   Light detection and ranging (LIDAR) via single-photon sensitive detector (SPAD) arrays is an emerging technology that enables the acquisition of depth images at high frame rates. However, the spatial resolution of this technology is typically low in comparison to the intensity images recorded by conventional cameras. To increase the native resolution of depth images from a SPAD camera, we develop a deep network built to take advantage of the multiple features that can be extracted from a camera's histogram data.  The network is designed for a SPAD camera operating in a dual-mode such that it captures alternate low resolution depth and high resolution intensity images at high frame rates, thus the system does not require any additional sensor to provide intensity images.  The network then uses the intensity images and multiple features extracted from down-sampled histograms to guide the up-sampling of the depth. Our network provides significant image resolution enhancement and image denoising across a wide range of signal-to-noise ratios and photon levels.  Additionally, we show that the network can be applied to other data types of SPAD data, demonstrating the generality of the algorithm. 
\end{abstract}
\maketitle


\section{Introduction}

\begin{figure*}
\centering
\includegraphics[width= 14 cm]{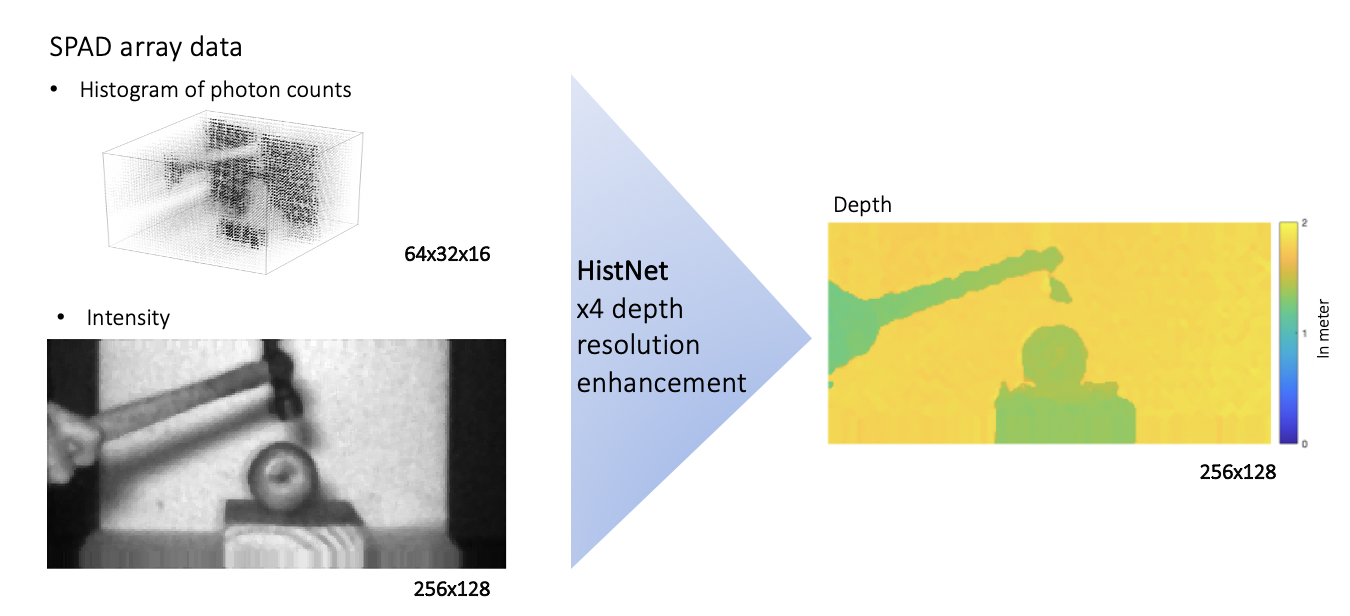}
\caption{\label{intro_figure} 
\textbf{Overview of the proposed network.} HistNet is designed for a SPAD camera operating in a dual-mode.  The camera provides a histogram of photon counts and an intensity image. The network takes this as an input and provides a HR depth map as the output with a resolution that is four times higher in both spatial dimensions than the initial raw histogram.}
\end{figure*}

Light detection and ranging (lidar), where a pulse of light is used to illuminate a target and a detector provides time-of-flight information, is one of the leading technologies for depth imaging.  For example, lidar is one of the key systems for future connected and autonomous vehicles, and it is used in the latest smartphones and tablet to aid auto-focus assist and enhance virtual reality.  Single-photon avalanche detector (SPAD) arrays are an emerging technology for depth estimation via lidar.  These are devices that are sensitive to single photons and can provide histograms of times of arrival of a single photons with respect to a trigger.  When used in combination with a pulsed laser that illuminates a target object, SPAD arrays provide accurate and fast data that can be converted to depth information. 

In the context of lidar, several different SPAD array sensors have been developed, see \cite{Ximenes2018, Zhang:18} for recent examples.  They have been used to measure depth in a range of scenarios, including under water \cite{Maccarone:15,7857033}, long range \cite{Chan2019, McCarthy:13, Pawlikowska:17}, at high speed \cite{Hutchings2019, 8662355, :GyongyOptica, Bronzi:14}, and providing high-resolution depth information \cite{Ren:18, Zhang:18}.  Recently, Morimoto \textit{et al.} reported a mega-pixel SPAD array \cite{Morimoto:20}.  SPAD array sensors have also been used for light-in-flight imaging \cite{Gariepy2015} and looking at objects hidden around corners \cite{Gariepy2016}.  They have also been used extensively within the field of biophotonics, see ref \cite{Bruschini2019} for a review.

Although SPAD arrays are becoming well established in lidar systems, there are several key challenges to overcome to fully exploit their potential.  The single-photon sensitivity that the SPAD array provides promises depth imaging at long ranges and in degraded visual environments, but improving the performance in these scenarios can dramatically increase the range of use of the detectors. In addition, the native resolution is typically very low in comparison to conventional image sensors.  Ultimately, it is desirable to operate the SPAD arrays at high frame rates, cover a large field-of-view at large distances, produce images at high resolutions, and perform well in a wide range of environmental conditions. Each of these objectives brings separate challenges that need to be addressed.

Due to the nature of the challenges to single-photon depth imaging, computational post-processing techniques are known to be a very powerful method to improve the overall image quality, both in terms of signal-to-noise and resolution. The latter methods \cite{:Tachella, :Rapp2017, Halimi-19, :DepthSR-Net, :Lindell-20, :Lindell} take advantage of prior information in one form or another and attempt to improve the quality of the depth images in the low-photon regime.  There are advantages and disadvantages to each of the methods, often with a trade off in terms of quality of the reconstruction and for the time taken to reconstruct.

Several statistical approaches have been implemented to improve the depth maps of single-photon depth data: Tachella \textit{et al.} established their images making use of priors on the depth and intensity estimates, achieving fast reconstruction and very good performance in the single-photon regime \cite{:Tachella}; Rapp and Goyal tackled the problem by the creation of super-pixels that borrow information from relevant neighbours to separate the contributions of signal and background  \cite{:Rapp2017}; and Halimi \textit{et al.} implemented an alternating direction method of multipliers (ADMM) to minimize a cost function using priors on correlations between pixels using both depth and intensity estimates \cite{Halimi-19}. Callenberg \textit{et al.} implemented an iterative optimization scheme to increase the spatial resolution of SPAD array data using a sensor fusion approach \cite{Callenberg21}. We refer the reader to \cite{Wallace:20,Rapp:20} for more details regarding state-of-the-art robust reconstruction algorithms.

Machine learning approaches have also shown good performance when enhancing the quality of high-resolution depth data.   For example, Guo \textit{et al} developed a deep neural network to reconstruct a high-resolution depth map from its low resolution version \cite{:DepthSR-Net}. In addition,  Lindell \textit{et al} \cite{:Lindell} and Sun \textit{et al} \cite{:Lindell-20} developed deep networks that process the whole 3D volume of raw photon counts and output a 2D depth map, achieving high performance in the low photon regime at the cost of a long processing time.  All references \cite{:DepthSR-Net, :Lindell, :Lindell-20} make use of an intensity image to guide the reconstruction of the depth.  The networks in references \cite{:Lindell, :Lindell-20} are specifically designed for SPAD array data, whereas the work in \cite{:DepthSR-Net} processes existing depth maps. Peng {\it et al.} \cite{Peng2020} implemented a network capable of reconstructing depth from SPADs at very long ranges by exploiting the non local correlations in time and space via the use of non local blocks and down-sampling operators. Recently Nishimura {\it et al.}~\cite{Nishimura2020} demonstrated how a single depth histogram can be used to resolve  the depth ambiguity problem associated with monocular depth estimation.

This paper proposes and implements a machine learning network to simultaneously perform up-sampling and de-noising of depth information.  We design and apply the network so that it is suitable for the data provided by the Quantic 4x4 sensor \cite{ Hutchings2019, 8662355, :GyongyOptica}, but the network can be applied to other depth images.  The SPAD camera alternates between two modes at over 1000 frames per second.  It provides high-resolution intensity images at a resolution of 256x128 pixels followed by low-resolution 64x32x16 histogram of photon counts containing depth information. 

Our approach is to select the essential information from the histogram and input this directly to the network, without requiring the entire histogram of counts to be provided to the network. After processing, the final resolution of our up-sampled depth images from the Quantic 4x4 sensor is increased by a factor of four to 256x128 pixels. Figure \ref{intro_figure} shows an overview of the results of the proposed network when applied on captured data.

This paper is organized as follows: in Section \ref{SPAD}, we provide a brief overview of the SPAD array sensor, the model of photon detection, and we present the processing done to the SPAD data to extract useful information prior to the reconstruction via the network. Section \ref{method} introduces the proposed HistNet in details. Section \ref{results} reports the results on both simulated and real data along with a comparison to other algorithms, and demonstrates its robustness to different noise scenarios.  Section \ref{conclusion} presents our conclusions and future work.

\section{SPAD array data}
\label{SPAD}
\subsection{Data acquisition}

Single-photon avalanche diode arrays can capture depth and intensity information of a scene. To achieve this, a short laser pulse is used to illuminate a target, and the detector records the arrival time of photons reflected back by the scene with respect to a laser trigger. This data, known as time tagged data, can be used to generate a temporal histogram of counts with respect to the time of flights, where the peak in the histogram can be used to calculate the distance to the target. 

In this work, we develop a network suitable for a SPAD array sensor, the Quantic 4x4 sensor, that generates a histograms of counts on-chip and operates in a hybrid acquisition mode \cite{ Hutchings2019, 8662355, :GyongyOptica}. This hybrid mode alternates between two measurement modes at a temporal rate exceeding 1000 frames per second. The details of the modes are: a high-resolution (HR) intensity measurement with a spatial resolution of 256x128, and a low-resolution (LR) histogram of photon counts containing depth information at a resolution of 64x32x16 (16 being the number of time bins of each of the 64x32 histograms).  It is the purpose of the network to increase the resolution of the initial depth data (64x32) to the same resolution as the intensity data (256x126), while simultaneously denoising the data.

\subsection{Pre-processing of data for the network}

\begin{figure*}
\centering
\includegraphics[width=12 cm]{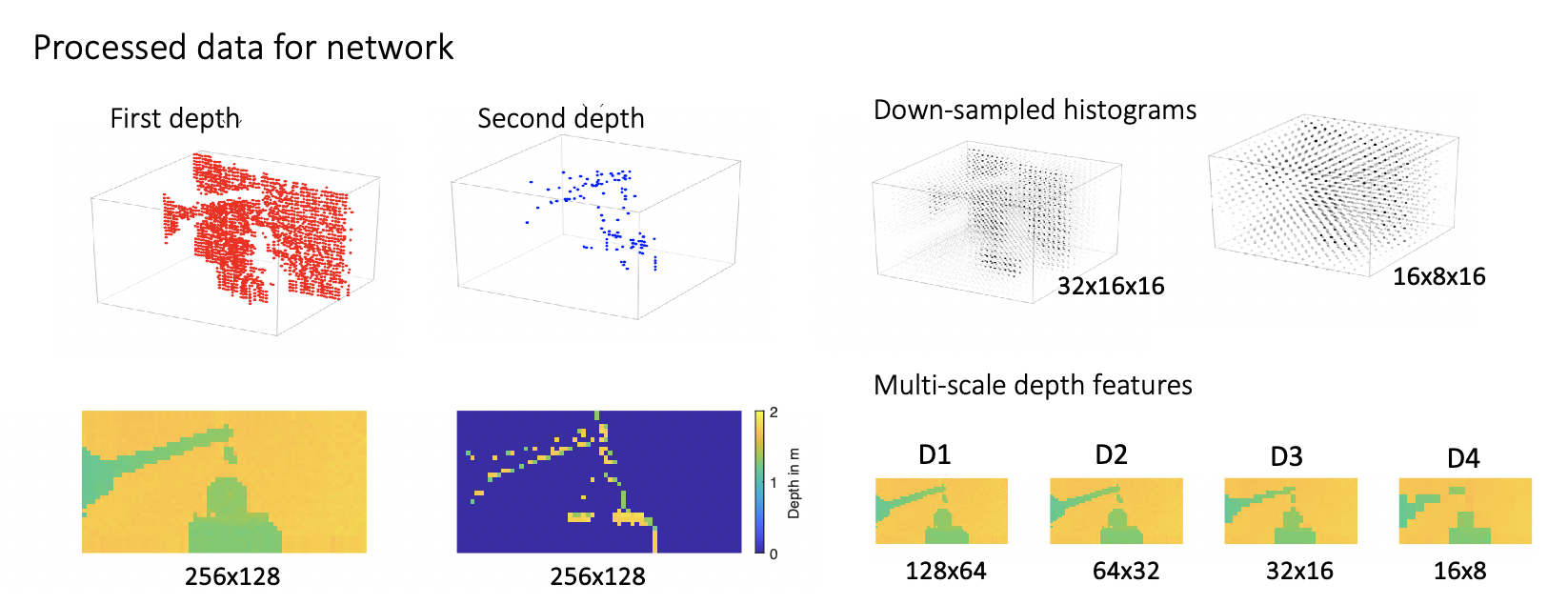}
\caption{\label{fig:DataProcessing}
\textbf{Representation of the processing of the different arguments of HistNet.} The SPAD array provides alternating LR Histograms of size 64x32x16 and HR intensity images of size 256x128. The 256x128 first and second depth maps are obtained by computing center of mass around the strongest and second strongest peaks in the raw histogram and then up-sampling by four in both spatial dimensions using nearest neighbour interpolation. Depth feature D1 of size 128x64 is obtained by down-sampling the first depth map by two in both dimensions. D2 of size 64x32 is obtained by applying center of mass around the strongest peak in the 64x32x16 raw histogram. D3 and D4 are obtained by down-sampling the raw 64x32 histogram respectively by two and four, and computing centre of masses around the strongest peak of subsequent down-sampled versions.}

\end{figure*}

The SPAD camera provides LR histogram data and HR intensity images in alternate frames. We will see in the following sections how we select features from the SPAD array data that maximise the quality of information provided to the network while minimising the total quantity of data necessary for accurate super-resolution. There are several different features that we extract from the data provided by the SPAD array: the first depth map, the second depth map, the high-resolution intensity image, and the multi-scale depth features extracted from down-sampled versions of the original histogram.  The processing time to calculate each of these features is minimal, adding very little computational overhead to our overall procedure.  Figure \ref{fig:DataProcessing} shows SPAD array data and the different processing steps to compute the arguments of HistNet. 

\subsubsection{First depth map}
\label{First depth map}

The first depth map is calculated directly from the 64x32x16 3D histogram data. The photon counts data can be assumed to be drawn from the Poisson distribution $\mathcal{P} \left(.\right)$ as commonly used in \cite{:Rapp2017, Halimi-19}. 
Assuming known background level and a Gaussian system impulse response, the maximum likelihood estimator of the depth is obtained as the central mass of the received signal photon time-of-flights (assuming depths are far from the observation window edges). This estimator is approximated for each depth pixel $(i,j)$ as 

\begin{equation}
\label{equation_center_of_mass}
d_{i,j} = \frac{\sum_{t = \max(1 , d^{max}-1)}^{\min(T ,  d^{max}+1)} t * \max(0 , h_{i,j,t}-b_{i,j})} {\sum_{t = \max(1 , d^{max}-1)}^{\min(T , d^{max}+1)} \max(0, h_{i,j,t}-b_{i,j})},
\end{equation}
with $h_{i,j,t}$ being the photon counts acquired in pixel $(i,j)$  for  time bin $t \in [1,T]$, $b_{i,j}$ is the background level of pixel $(i,j)$ estimated as the median of each pixel, and $d^{max}$ represents the location of the signal peak estimated as the location of the  maximum of the histogram of counts in photon dense regimes, or using a matched filter in sparse photon regimes. We integrate over three time bins, between $d^{max}-1$ and  $d^{max}+1$, as this width corresponds approximately to the impulse function of our system. Before this data is input to the network, we divide the depth by the total number of time bins to normalize it between 0 and 1.  It is then up-scaled to the size of the desired resolution (four times larger in both spatial dimensions) using a nearest neighbour interpolation. Nearest neighbour interpolation is used as it preserves the separate surfaces in the scene.  This is preferable over other interpolation strategies that connect separate surfaces with new depths that did not exist in the original data.

\subsubsection{Multi-scale depth features}

Multiple resolution scales have been shown to help depth estimation, in particular in high noise scenarios \cite{:GyongyOptica, :Rapp2017, :DepthSR-Net}. This information is included to the network by using four depth features   D1, D2, D3 and D4 of different resolution scales. The dimensions of each feature for our real data (64x32x16 histogram) are 128x64, 64x32, 32x16 and 16x8 respectively.
D1 is obtained by down-sampling the previously obtained 256x128 first depth map by two in both spatial dimensions using nearest neighbour interpolation, by picking only the top-left pixel within squares of 2x2 pixels. D2 is obtained by computing center of mass on the 64x32x16 LR histogram. D3 and D4 are obtained by down-sampling this histogram by a factor of two and four by summing the neighbouring pixels in the spatial dimensions, hence obtaining histograms of size 32x16x16 and 16x8x16 respectively. D3 and D4 are then obtained by computing center of mass on those histograms. Thanks to this process of down-sampling at the level of histogram, the resultant D3 and D4 have higher signal-to-noise than the first depth map, albeit with a lower resolution. This helps the network identify features in images with high levels of noise. All the features are normalized by dividing with the total number of time bins of the system (i.e., the width of the range observation window).

\subsubsection{High resolution intensity}
The intensity map of a scene has been used to guide the reconstruction of the depth in statistical methods \cite{:Tachella, :GyongyOptica, :Rapp2017, Halimi-19} and in machine learning methods \cite{:Lindell, :DepthSR-Net, :Lindell-20}.  In our case, we obtain the intensity image directly from the SPAD detector Quantic 4x4. This intensity image has a spatial resolution of 256x128, which is four times larger than the 64*32 histogram spatial resolution. We normalize between 0 and 1 this intensity before inputting to the network. The intensity and histogram data are acquired in alternate frames, so it is possible that objects can move from frame to frame. However, our system has a high temporal frame rate, so we assume perfect alignment between the histogram and intensity data.

\subsubsection{Second depth map}
When the Quantic 4x4 sensor operates in the depth mode, each superpixel in the histogram gathers the photon counts of 4x4 pixels. Therefore, some histograms might present multiple peaks when observing multiple surfaces located at different depths. Those histograms contain multiple peaks that correspond to the different depths involved. While the first depth map is calculated by identifying the strongest peak, we compute a second depth map based on the second strongest peak. More precisely, the depth position in the second depth map is calculated by applying the center of mass of Equation \ref{equation_center_of_mass} around the index of the second strongest peak. 

We set the following criterion on the minimum number of photon counts a relevant peak should contain. For each pixel $(i,j)$ we consider a peak at bin $t$ to be relevant if
\begin{equation}
\label{criterion}
   h_{i,j,t} >  b_{i,j} + level*\sqrt{b_{i,j}}
\end{equation}
with $h_{i,j,t}$ being the number photon counts of pixel $(i,j)$ at time bin $t$; $b_{i,j}$ being the background level at pixel $(i,j)$ estimated by taking the median value of the histogram; $\sqrt{b_{i,j}}$ represents the standard-deviation of the Poisson distributed background counts; and $level$ is a variable adjusted empirically, so that the values in the second depth map do not come from the noise but mostly represent real signal. In the case of our captured data, we set $level$ to 12.
If the criteria is not met, the second depth is set to zero.

We note that in high noise scenarios, discriminating peaks corresponding to real depths from peaks corresponding to background photons is difficult. In extremely noisy scenarios, no second depth can be extracted and the second depth map is  set to zero. In the same way as for the first depth map, the second depth map is 4x4 up-scaled with a nearest neighbour interpolation.  
The second depth map can be computed in the case of the Quantic 4x4 camera because this sensor gathers the photon counts of 4x4 pixels in the depth mode. In the case of more traditional SPAD sensors, e.g. in section 4.\ref{8times}, the second depth map is set to zero.

\subsubsection{Outlier rejection step}
\label{Outlier_rejection_step}
For the cases when the objects are contained only within a few bins, e.g.~in section 4.\ref{8times}, we crop the LR histogram in the temporal dimension, so that the returned signal from the studied objects extends over all time bins. In  that case, all the previously detailed depth features (i.e. first depth map, second depth map, and the four depth features D1, D2, D3 and D4) are computed on the cropped LR histogram. This step ensures the maximum contrast of the depth features and the best performance of the network. We evaluate the range of time bins in which the objects are present in the following way. For one scene, the histograms of all pixels are added into one single histogram. On this single histogram, we use criterion \eqref{criterion} (to estimate the second depth map), to select the bins where a minimum photon count is received, and then apply a median filter to erase all isolated peaks. The range of the remaining time bins is used to crop the LR histogram in the temporal dimension. 

\section{Proposed method}
\label{method}
The proposed HistNet network increases the spatial resolution of the input depth map by four in both dimensions (e.g., from 64x32 to 128x64 for our real data). It is also robust to extremely noisy conditions as highlighted in following sections. The network structure is independent of the input dimensions and can therefore reconstruct data of any size. However, for clarity, we use the dimensions of our real data to present the network structure.

\subsection{Network architecture}
\label{Network architecture}

 In the context of guided depth super-resolution, Guo \textit{et al.} \cite{:DepthSR-Net} developed the U-Net-based \cite{:UNet} DepthSR-Net algorithm, which offers state-of-the-art performance.  The U-Net architecture is particularly useful as it requires very few training data and is based on CNNs which have shown good results for image denoising and super-resolution tasks \cite{:Lindell, :Lindell-20}.

\begin{figure*}
\centering
\includegraphics[width= 14 cm]{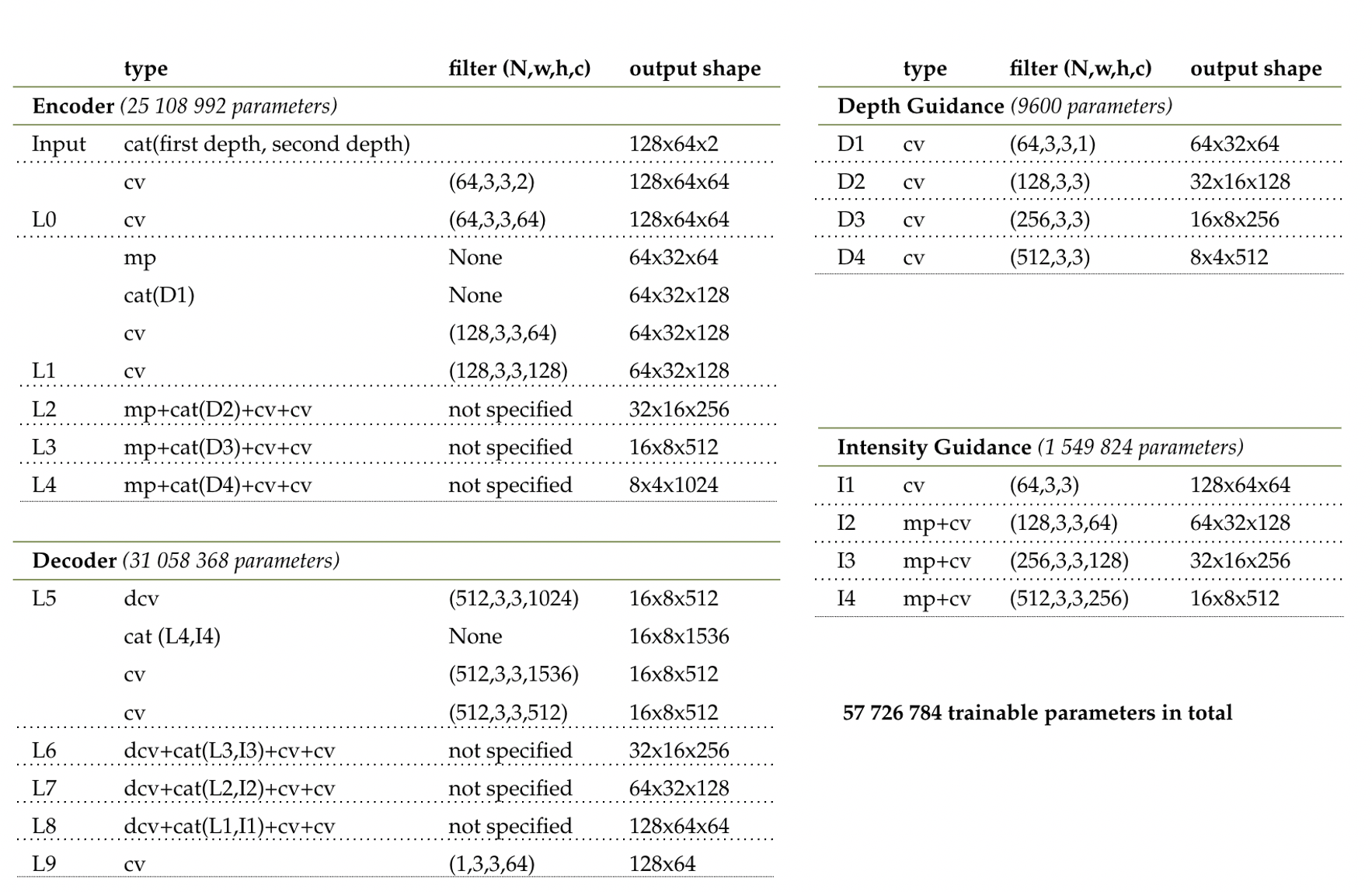}
\caption{\label{details_network} 
\textbf{Details of network architecture and associated parameters}. The  "output shapes" of the layers specified in the fourth column correspond to the case of processing our real data (i.e. histograms of spatial resolution 64x32 and 128x64 intensity images). The filters of each layer are described with four parameters: N stands for the number of filters; w the width; h the height and c the number of inner channels. cv stands for convolutional layer; mp for max-pooling layer; cat() for concatenation with the layers specified in the brackets and dcv for deconvolutional layer.}
\end{figure*}

\begin{figure*}
\centering
\includegraphics[width= 14 cm]{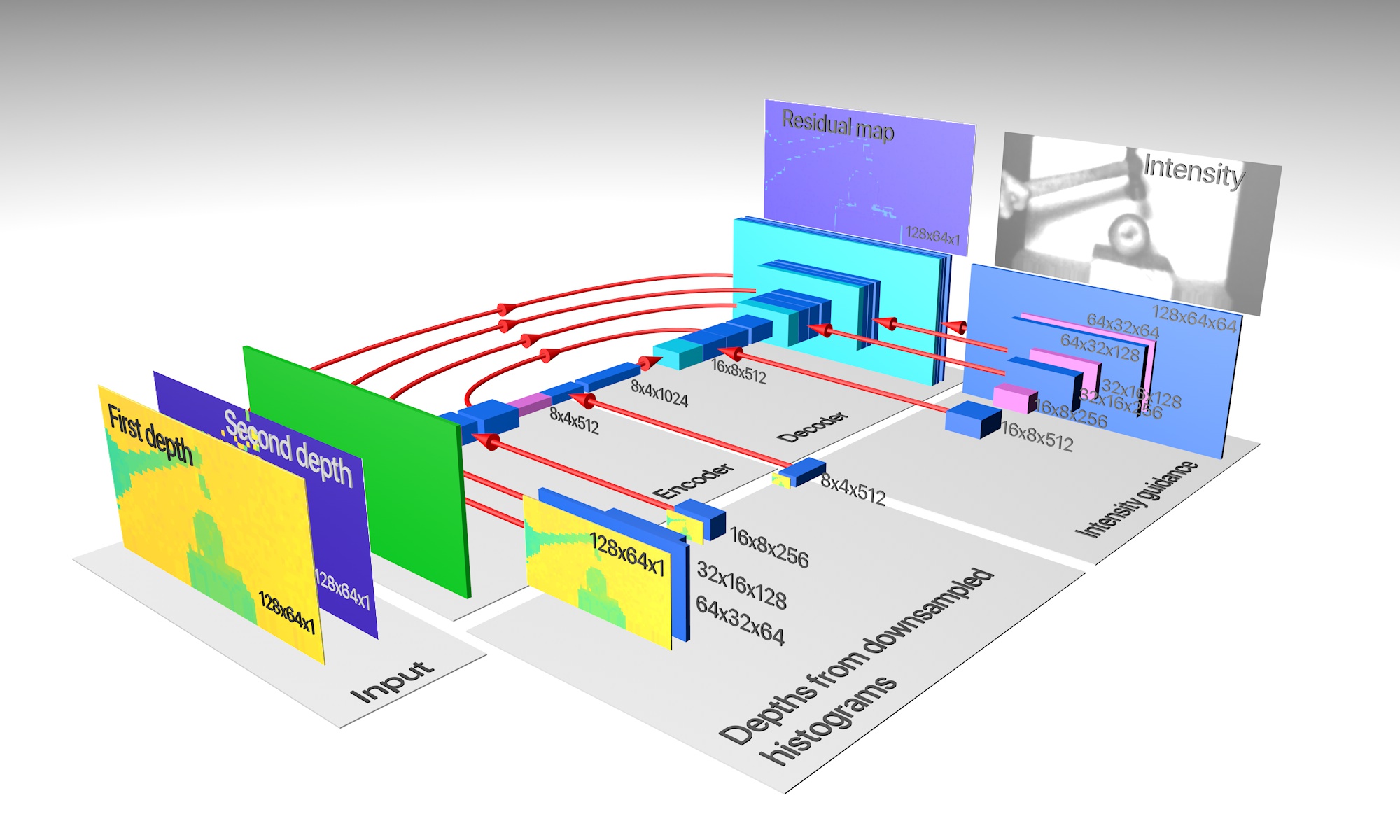}
\caption{\label{fig:Network}
\textbf{Representation of the HistNet.} Input of the network consists of the first and the second depth maps. Multi-resolution depths features are integrated along the contracting path of the U-Net. The intensity image is processed at multiple resolution and integrated along the expansive part of the U-Net. Skip connections between the contracting and the expansive paths are displayed as red arrows.}
\end{figure*}

In this paper, we develop a new network to account for the multi-scale information available from the SPAD array histogram data, making it more robust to sparse photon regimes and/or high background scenarios, in addition to exploiting the fine details of the intensity guide.  The network makes use of the different features extracted from the raw histogram (first and second depth maps, the multi-scale depths) and the intensity image. Our network performs simultaneous up-sampling and denoising of depth data for a wide range of scenarios. Details of the number and filter sizes for each layer are reported in Figure \ref{details_network}. Figure \ref{fig:Network} shows a schematic representation of the network. 

With respect to other state-of-the art algorithms, HistNet differs in that it extracts the key information from a histogram prior to the data being processed by any neural network.  This provides the network with known informative features, allowing a reduction in processing times.

\subsubsection{Residual U-Net architecture}
\label{architecture_Net}

The goal of the network is to take the data from the low-resolution histogram and intensity image and produce a residual map $\mathcal{R}$ that can be added to an up-scaled version of the low resolution depth map \cite{he2015deep, :DepthSR-Net}.  The sum of the residual map and the low-resolution depth map is the final high-resolution depth map.  The goal of the training is to find the parameters of the filters, i.e.~the weights and biases, that minimises the $l1$-norm between the residual map and the ground truth.  

The network consists of an encoder of five layers, denoted L0 to L4, connected to a five-layer decoder (L5 to L9) with skip connections.  The network includes a branch that processes the multi-scale depth features (see \ref{multi_scale_dep}) and a branch that processes the intensity image (see \ref{multi_scale_int}). The main input of the encoder is the concatenation of the first and second depth maps along a third dimension. In the case of the real data, this input is therefore of size 128x64x2.  Note that each filter of the convolutional and deconvolutional layer has a height and width of 3.

In the encoder, the main input is passed to layer L0, which consists of two convolution operations of 64 filters each. Layers L1 to L4 consist of three steps: first, 2x2 max-pooling that down-samples the data by two in both spatial dimensions; second, integration of the information of the multi-scale depth features by concatenation with the layer of the depth guidance that has the same shape (see \ref{multi_scale_dep}); and finally, two convolution operations with a number of filters of 128, 256, 512 and 1024 for L1, L2, L3 and L4, respectively.

In the decoder, the layers L5 to L8 consist of four steps: first, deconvolution operations filter and up-sample the image by two in both spatial dimensions; second, skip connections between encoder and decoder are computed by concatenating the decoder layer with the layer in the encoder that has the same shape; third, guidance with the intensity is provided by concatenation with the right layer in the intensity guidance branch (see \ref{multi_scale_int}); and finally, two convolution operations are performed with a number of filters of 512, 256, 128 and 64 for L5 to L8, respectively.
L9 is a convolutional layer with one filter that provides the output.  The predicted HR depth map is obtained by adding the output of the network $\mathcal{R}$, which is known as the residual map, to the LR depth input, i.e. the first depth map.

\subsubsection{Guidance with the noise-tolerant multi-scale depth features}
\label{multi_scale_dep}
The multi-scale depth features are of size 64x32, 32x16, 16x8 and 8x4. Each feature passes through a convolutional layer of 64, 128, 256 and 512 filters of size 3x3 respectively, creating cubes of size 64x32x64, 32x16x128, 16x8x256 and 8x4x512. These cubes are integrated in the encoder by concatenation along the filter dimension with the layer of corresponding size (see \ref{architecture_Net}). 

\subsubsection{Guidance with the intensity map}
\label{multi_scale_int}
We use intensity guidance in the same manner as Guo \textit{et al.}~\cite{:DepthSR-Net}.
The guidance branch consists of convolutional operations followed by 2x2 max-pooling. The number of filters of the convolution operation of each layer is 64, 128, 256 and 512. For our real data, the outputs of the convolutional operations of each layer in the guidance branch are of size 128x64x64, 64x32x128, 32x16x256 and 16x8x512. These outputs are integrated along the decoder part of the network by concatenation along the filter dimension with the layer of corresponding size (see \ref{architecture_Net}).

\subsection{Loss}

We minimize the $l1$ loss to help reconstruct separate surfaces in depth. The $l1$-norm is known to promote sparsity  \cite{Lasso, l1_norm}. During the training, a batch-mode learning method with a batch size of $M$ = 64 was used, and the loss was defined as
\begin{equation}
    \mathcal{L}(\theta) = \dfrac{1}{MN} \sum_{m=1}^{M} \sum_{n=1}^{N} \left |{ \mathcal{R}_{m,n}(\theta) + d_{m,n} - d^{ref}_{m,n}} \right|,
\end{equation}
with $M$ the number of images within one batch, $N$ the number of pixels of each image, $\theta$ the trainable parameters of the network, the $\mathcal{R}$ the residual map predicted by HistNet, $d$ the first depth map, and $d^{ref}$ the ground truth depth.

\subsection{Simulated datasets for training, validation and testing}
\label{Training-Dataset}
We simulate realistic SPAD array measurements (LR histogram and HR intensity) from 23 scenes of the MPI Sintel Depth Dataset \cite{Butler:ECCV:2012, Wulff:ECCVws:2012} for the training and validation dataset, and from six scenes of the Middlebury dataset \cite{:Midd1,:Midd2} for the test dataset. 

From the HR depth and intensity images provided by both datasets, we create histograms of counts of 16 time bins. Here we simulate the case where the return signal from objects extends over all time bins of the histogram, and therefore no outlier rejection step is necessary (see \ref{Outlier_rejection_step}). The photon counts of the histogram are assumed to be taken from a Poisson distribution and the impulse response of our SPAD camera is approximated by a Gaussian function $\mathcal{G}(m, \sigma)$ with average m and standard deviation $\sigma=0.5714$ histogram bins \cite{8662355, :GyongyOptica}. For each pixel $(i,j)$, the photon counts $h_{i,j,t}$ of the acquired histogram at time bin $t$ can be expressed as function of the intensity $r_{i,j}$ and the depth $d_{i,j}$ as follows 
\begin{equation}
    h_{i,j,t} =\mathcal{P}  \left( r_{i,j} * \mathcal{G} (d_{i,j} , \sigma) + b_{i,j,t} \right),
\end{equation}
with $b_{i,j,t}$ the background level which is assumed constant for all time bins of a given pixel. To reconstruct the real measurements, we estimate the background $b_{i,j,t}$ of the simulated training dataset as the median of the histograms of the real measurements. For the simulated measurements, the background $b_{i,j,t}$ of the simulated training dataset is estimated from the SBR value. The LR histograms are simulated by down-sampling the HR histogram by integrating over four by four pixels in the spatial dimensions. We consider two metrics to assess the level of noise: the signal to background ratio (SBR) and the number photon counts per pixel ($ppp$). We define the pixel $SBR_{i,j}$ as
\begin{equation}
\label{equation_SBR}
SBR_{i,j} = \frac{\sum_{t = \max(1 , d^{max}-1)}^{\min(T ,  d^{max}+1)}  (h_{i,j,t}-b_{i,j})} {b_{i,j}*(\min(T ,  d^{max}+1) - \max(1 , d^{max}-1))}.
\end{equation}
The image SBR is then the average value over all pixels, i.e.~$SBR = \frac{1}{N} \sum_{i,j}{SBR_{i,j}}$. In a similar fashion, we define $ppp_{i,j}$, the number of photon counts reflected from the target in a pixel (i,j), as
\begin{equation}
\label{equation_ppp}
ppp_{i,j} = \sum_{t = \max(1 , d^{max}-1)}^{\min(T ,  d^{max}+1)}  h_{i,j,t}-b_{i,j}
\end{equation}
The average $ppp$ for the image is then the average value over all the pixels, i.e.~$ppp = \frac{1}{N} \sum_{i,j}{ppp_{i,j}}$.

It should be noted that the 436x1024 MPI Sintel Depth Dataset are pre-processed to remove all NaN and Inf values using median filtering.  We use 21 images for training and two
are saved for the validation. We increase the number of training images by a factor of eight by using all possible combinations of 90° image rotations and flips. Furthermore, the images are split into overlapping patches of size 96x96 with a stride of 48.

\subsection{Implementation details}

We implemented HistNet within the Tensorflow framework and use the  ProximalAdagradOptimizer optimizer \cite{AdagradProximalOpt}, as this enables the minimization of the $l1$ loss function. The learning rate was set to 1e-1. The training was performed on a NVIDIA RTX 6000 GPU. We trained for 2000 epochs, which took about 10 hours.

\newpage

\section{Results}
\label{results}

\subsection{Evaluation on simulated measurements (4-times up-sampling)}
\subsubsection{Noise scenarios}
Three different noise scenarios were considered: a scenario mimicking the lighting conditions of \cite{:GyongyOptica} with ppp = 1200 counts and SBR = 2 denoted "high signal-to-noise scenario", a scenario corresponding to a lower photon count and lower signal to noise with ppp = 4 counts and SBR = 0.02 denoted "medium signal-to-noise scenario", and a scenario corresponding to a lower photon count and much lower signal to noise with ppp = 4 counts and SBR = 0.006 denoted "low signal-to-noise scenario" . We trained a separate network for the three different noise scenarios.

\subsubsection{Evaluation metrics}
We quantify the performance of the network on simulated data by comparing the reconstruction with the ground truth depth using the root mean squared error metric RMSE $ = \sqrt { \frac{1}{N} \|{\mathcal{R}+d-d^{ref}}\|^2} $ and the absolute depth error ADE $= \left|{\mathcal{R}+d-d^{ref}}\right|  $, $\mathcal{R}$ being the residual map predicted by HistNet after the training, $d$  the up-scaled version of the low resolution depth map, i.e the first depth map, and $d^{ref}$ the ground truth.

\subsubsection{Comparison algorithms}
We compare the results of HistNet with the following methods:

\begin{itemize}
    \item \textit{Nearest-neighbour interpolation}  Depth is estimated with center of mass on the 276x344x16 histogram of counts and up-sampled with nearest neighbour interpolation  to a 1104x1376 image. Note that this is the method used to produce the first depth argument of HistNet. We choose nearest-neighbour interpolation to avoid joining spatially separated surfaces.  
    
    \item \textit{Guided Image Filtering of He \textit{et al.}~2013} \cite{guidedfiltering} We perform further processing to the estimated depth from the nearest-neighbour interpolation by applying the guided filtering algorithm with the HR intensity image as a guide.
    
    \item \textit{DepthSR-Net of Guo \textit{et al.}~2019} \cite{:DepthSR-Net}  We retrained this network using the same training datasets for our network. This network outputs a 4x upsampled depth map from a LR depth map using a HR intensity map to guide the reconstruction. 
    In \cite{:DepthSR-Net}, the LR depth map is first up-sampled to the desired size with a bicubic interpolation. However, we want to reconstruct surfaces that are well separated from one another. Therefore, nearest neighbour interpolation instead of bicubic interpolation is used to upsample the input LR depth map. 
    
    \item \textit{Algorithm of Gyongy et al.~2020} \cite{:GyongyOptica} This  algorithm is designed to process the Quantic 4 x4 SPAD array.  It consists of various steps of guided filtering and up-sampling with low computational cost. One part of the algorithm is designed to compensate the inherent misalignement between the depth and the intensity information that the SPAD provides.  Since our synthetic data consists of perfectly aligned intensity and depth, we do not use this part of the algorithm. 
\end{itemize}

\subsubsection{Results for the different signal-to-noise scenarios}
\begin{figure*}
\includegraphics[width=\textwidth]{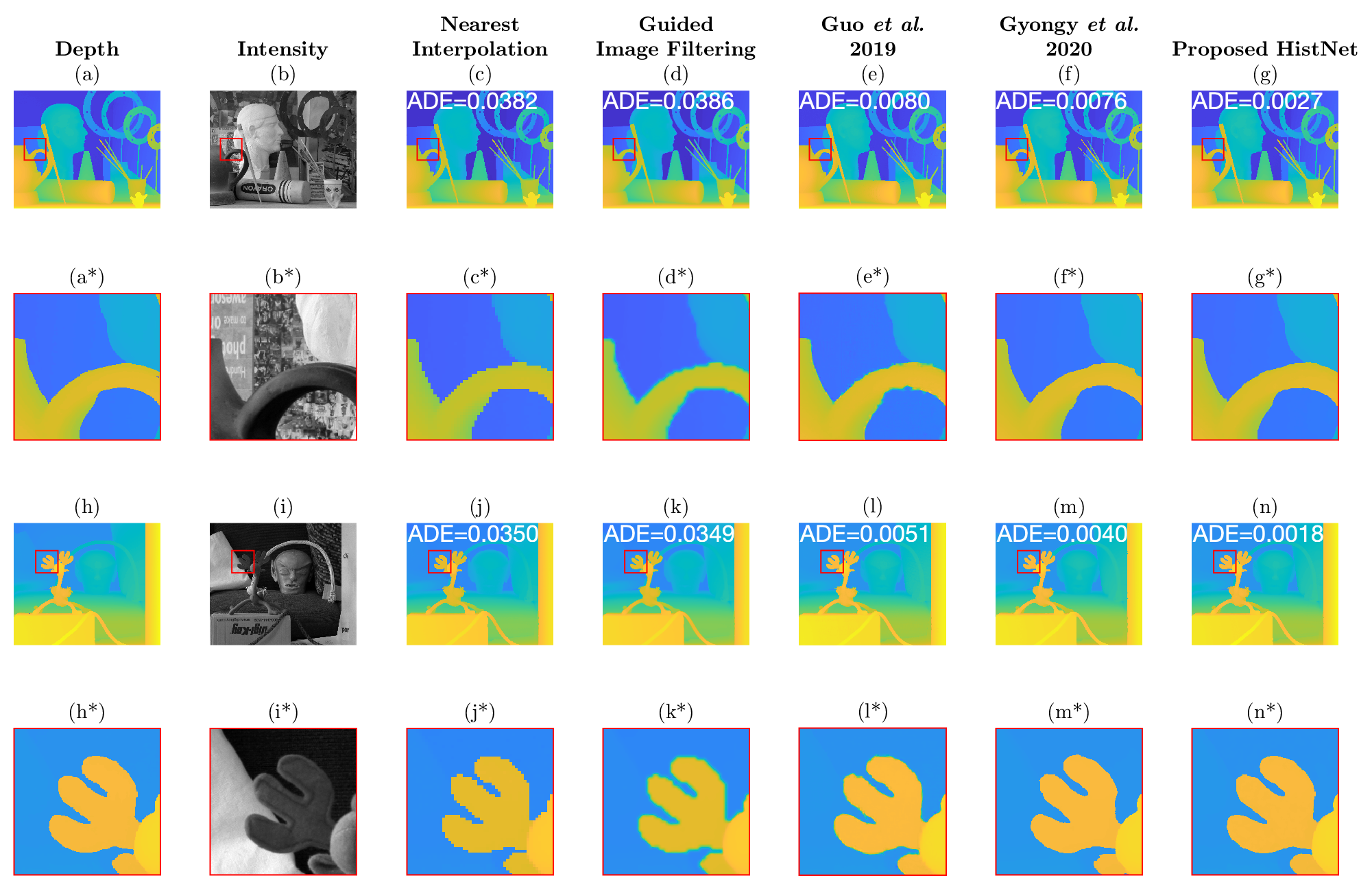}
\caption{\label{fig:ppp1200_SBR2} \textbf{High signal-to-noise. Comparison of reconstruction techniques for measurements simulated with an average of 1200 signal photons per pixel and a signal-to-background ratio of 2.} ADE is the absolute depth error calculated with normalized data between 0 and 1. (a) displays the ground truth depth image; (b) displays the ground truth of the intensity; (c) is the reconstruction of the depth with nearest interpolation; (d) is the up-sampled version of the first depth data with Guided Image Filtering \cite{guidedfiltering}; (e) is the reconstruction of the first depth data with the algorithm DepthSR-Net of Guo \textit{et al.} 2019 \cite{:DepthSR-Net}; (f) is the reconstruction of the first depth data with the algorithm of Gyongy \textit{et al.} 2020 \cite{:GyongyOptica}; (g) is the reconstruction via our proposed method HistNet.  (a*), (b*), (c*), (d*), (e*), (f*) and (g*) are the closeup views of (a)-(g). (h)-(n) are the corresponding images from another simulated measurements with (h*)-(n*) the closeup views.
}
\end{figure*}

Figure \ref{fig:ppp1200_SBR2} shows the reconstruction for the different reconstruction methods for high signal-to-noise scenario: (ppp = 1200 counts and SBR = 2). We see that HistNet is able to produce sharp and clean boundaries. Guided Filtering and DepthSR-Net introduce blurred details around the edges and nearest interpolation leads to pixelated images. We report the root mean squared error and the absolute error in Table \ref{table:compa4x} for the two Middlebury scenes reconstructed with the different methods. This table indicates that HistNet outperforms the other methods in both performance categories. The processing time of the different methods is reported in Table \ref{table:compa4x}. Guided Filtering and nearest interpolation have a very low computational cost and process the image the fastest, in a few milliseconds for a 1104x1376 input. The algorithm of Gyongy \textit{et al.} \cite{:GyongyOptica} reconstructs the image in about 4 seconds. The reconstructions of HistNet and DepthSR-Net were performed on a NVIDIA RTX 6000 GPU. Each 1104x1376 Middlebury scene took about 7 seconds to reconstruct with the two networks.

\begin{figure*}
\includegraphics[width=\textwidth]{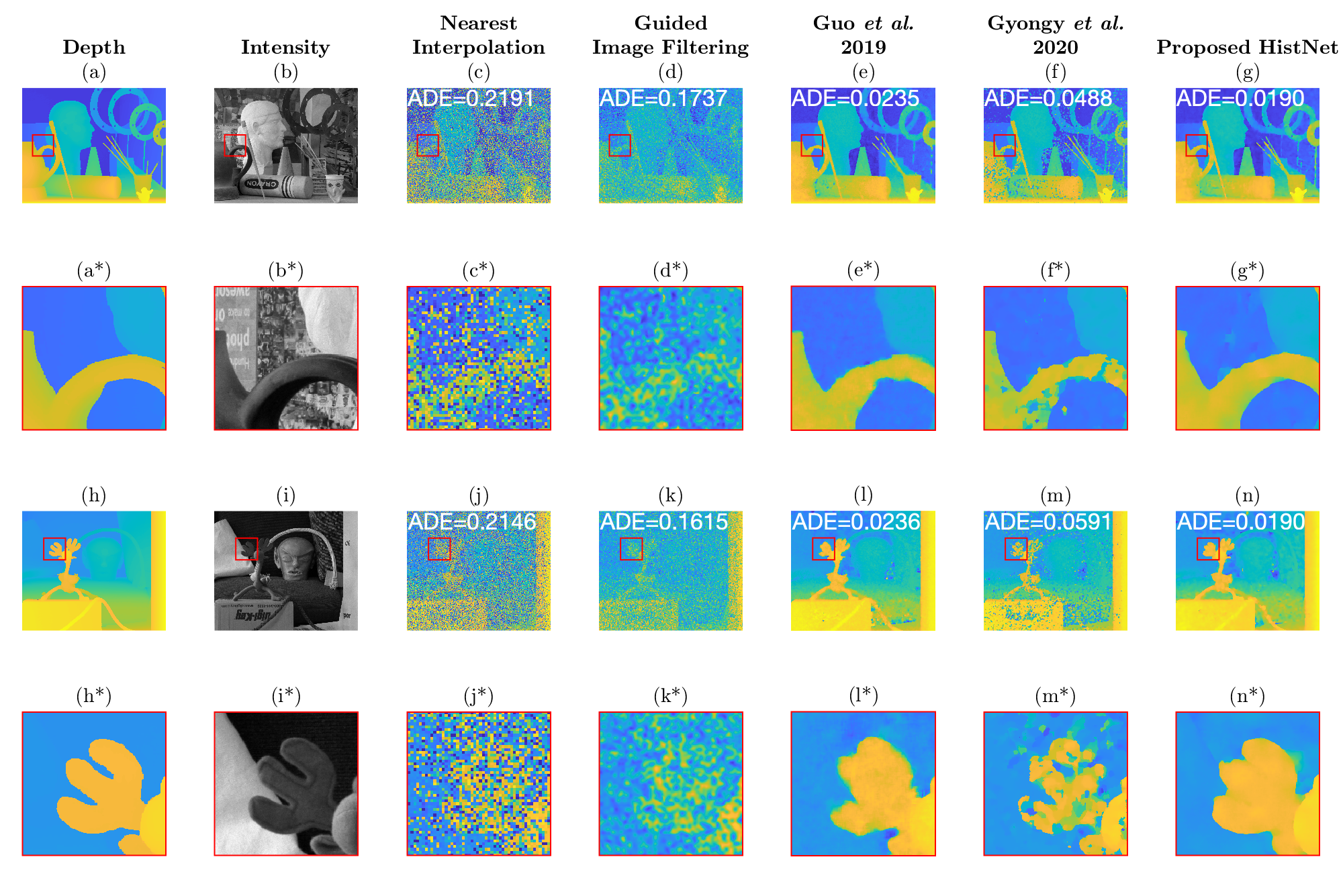}
\caption{\label{fig:ppp4_SBR0.02} \textbf{Medium signal-to-noise. Comparison of reconstruction techniques for measurements simulated with an average of 4 signal photons per pixel and a signal-to-background ratio of 0.02.}  ADE is the absolute depth error calculated with normalized data between 0 and 1. (a) displays the ground truth depth image; (b) displays the ground truth of the intensity; (c) is the reconstruction of the depth with nearest interpolation ; (d) is the up-sampled version of the first depth data with Guided Image Filtering \cite{guidedfiltering}; (e) is the reconstruction of the first depth data with the algorithm DepthSR-Net of Guo \textit{et al.} 2019 \cite{:DepthSR-Net}; (f) is the reconstruction of the first depth data with the algorithm of Gyongy \textit{et al.} 2020 \cite{:GyongyOptica}; (g) is the reconstruction via our proposed method HistNet.  (a*), (b*), (c*), (d*), (e*), (f*) and (g*) are the closeup views of (a)-(g). (h)-(n) are the corresponding images from another simulated measurements with (h*)-(n*) the closeup views.}
\end{figure*}

\begin{figure*}
\includegraphics[width=\textwidth]{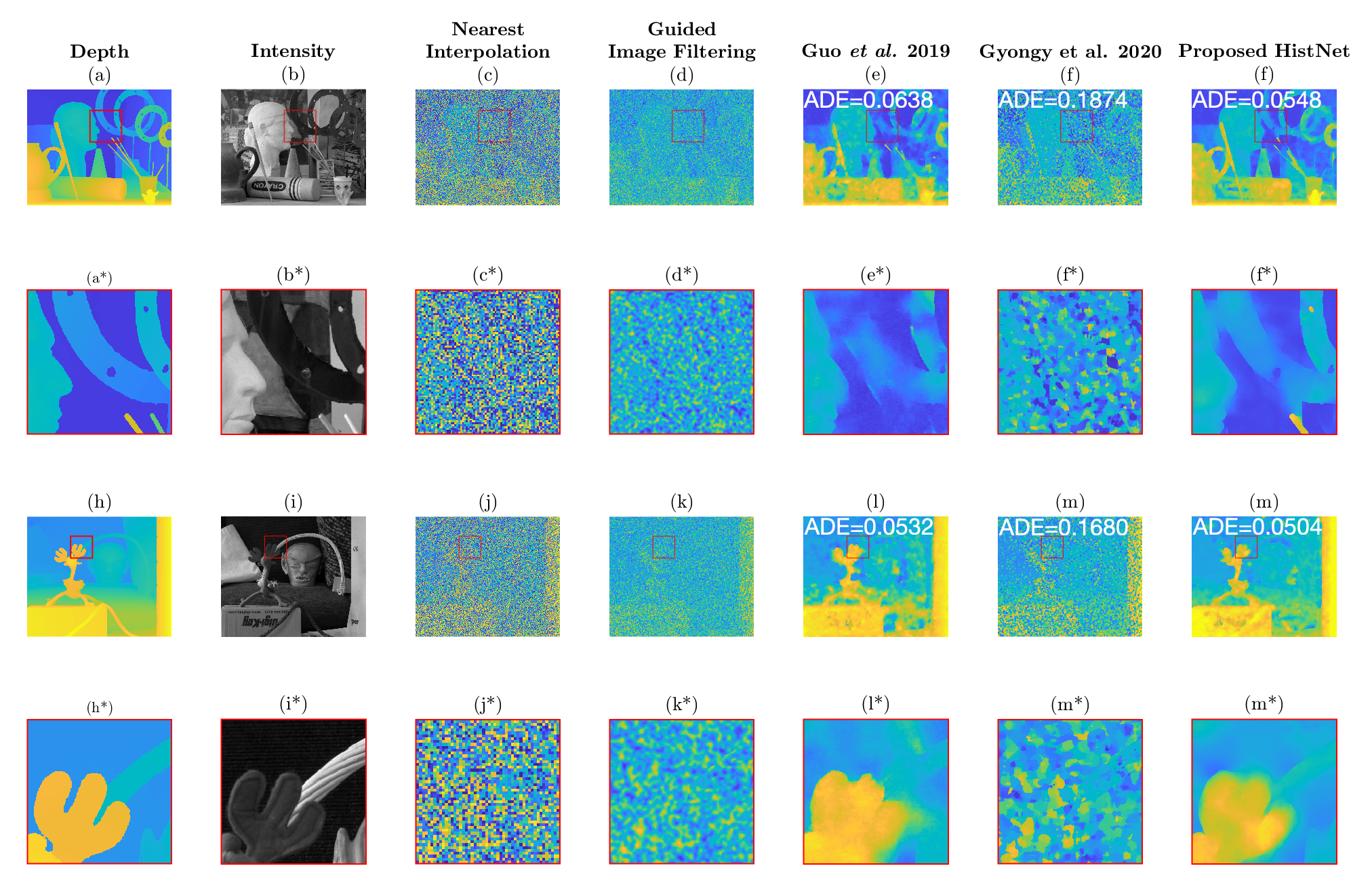}
\caption{\label{fig:ppp4_SBR0_006} \textbf{Low signal-to-noise. Comparison of reconstruction techniques for measurements simulated with an average of 4 signal photons per pixel and a signal-to-background ratio of 0.006.} ADE is the absolute depth error calculated with normalized data between 0 and 1. (a) displays the ground truth depth image; (b) displays the ground truth of the intensity; (c) is the reconstruction of the depth with nearest interpolation; (d) is the up-sampled version of the first depth data with Guided Image Filtering \cite{guidedfiltering}; (e) is the reconstruction of the first depth data with the algorithm DepthSR-Net of Guo \textit{et al.} 2019 \cite{:DepthSR-Net}; (f) is the reconstruction of the first depth data with the algorithm of Gyongy \textit{et al.} 2020 \cite{:GyongyOptica}; (g) is the reconstruction via our proposed method HistNet.  (a*), (b*), (c*), (d*), (e*), (f*) and (g*) are the closeup views of (a)-(g). (h)-(n) are the corresponding images from another simulated measurements with (h*)-(n*) the closeup views.
}
\end{figure*}

Figure \ref{fig:ppp4_SBR0.02} and \ref{fig:ppp4_SBR0_006} show the results for measurements simulated with an medium signal-to-noise (ppp=4 counts, SBR =0.02) and low signal-to-noise (ppp=4 counts, SBR =0.006), respectively.  Visually, we see that our method performs the best in high noise scenarios. Quantitative comparison can be found in Table \ref{table:compa4x}. For both scenes, HistNet performs better in terms of the RMSE and ADE.



\begin {table}[!ht]

\footnotesize
\begin{center}
\begin{tabular}{|m{2cm}|m{1cm}|m{1cm}|m{1cm}|m{1cm}|m{1cm}|m{1cm}|m{1cm}|m{1cm}|m{1cm}|m{1cm}|} 
\hline
 & \multicolumn{2}{c|}{\textbf{Proposed HistNet}} & \multicolumn{2}{c|}{\textbf{Gyongy \textit{et al.}} \cite{:GyongyOptica}} & \multicolumn{2}{c|}{\textbf{Guo \textit{et al.}} \cite{:DepthSR-Net}} & \multicolumn{2}{c|}{\textbf{NNI}} & \multicolumn{2}{c|}{\textbf{Guided Image Filtering}} \\ \hline
Rec time per scene  & \multicolumn{2}{c|}{7s (on GPU)} & \multicolumn{2}{c|}{4s} & \multicolumn{2}{c|}{7s (on GPU)} & \multicolumn{2}{c|}{1ms} & \multicolumn{2}{c|}{0.4s} \\ \hline
 \multicolumn{11}{|c|}{Training on high signal-to-noise data with second depth; ppp=1200 counts and SBR=2} \\ \hline
\textbf{Scene} & \textbf{RMSE} & \textbf{ADE} & \textbf{RMSE} & \textbf{ADE} & \textbf{RMSE} & \textbf{ADE} &\textbf{RMSE} & \textbf{ADE} & \textbf{RMSE} & \textbf{ADE}\\ \hline
Art & \textbf{0.023} &  \textbf{0.0027} & 0.043 & 0.0076 & 0.026 & 0.0080 & 0.053 & 0.038 & 0.046 & 0.039 \\ \hline
Reindeer &  \textbf{0.012} & \textbf{0.0018} & 0.023 & 0.0040 & 0.015 & 0.0051 & 0.040 & 0.035 & 0.037 & 0.035 \\ \hline
 \multicolumn{11}{|c|}{Training on medium signal-to-noise data without second depth; ppp=4 counts and SBR=0.02} \\ \hline
 \textbf{Scene} & \textbf{RMSE} & \textbf{ADE} & \textbf{RMSE} & \textbf{ADE} & \textbf{RMSE} & \textbf{ADE} &\textbf{RMSE} & \textbf{ADE} & \textbf{RMSE} & \textbf{ADE}\\ \hline
Art & \bf{0.053} &  \bf{0.019} & 0.11 & 0.050 & 0.054 & 0.023 & 0.32 & 0.22 & 0.22 & 0.17 \\ \hline
Reindeer &  \bf{0.040} & \bf{0.019} & 0.12 & 0.060 & 0.047 & 0.024 & 0.31  & 0.21 & 0.21  & 0.16 \\ \hline
 \multicolumn{11}{|c|}{Training on low signal-to-noise data without second depth; ppp=4 counts and SBR=0.006} \\ \hline
 \textbf{Scene} & \textbf{RMSE} & \textbf{ADE} & \textbf{RMSE} & \textbf{ADE} & \textbf{RMSE} & \textbf{ADE} &\textbf{RMSE} & \textbf{ADE} & \textbf{RMSE} & \textbf{ADE}\\ \hline
Art & \bf{0.082} & \bf{0.055} & 0.248 & 0.187 & 0.102 & 0.064 & 0.363 & 0.276 & 0.27 & 0.22 \\ \hline
Reindeer &  \bf{0.075} & \bf{0.050} & 0.234 & 0.168 & 0.083 & 0.053 & 0.357  & 0.272 & 0.259 & 0.206 \\ \hline
\end{tabular}
\caption{\label{table:compa4x} \textbf{Quantitative comparison of the different reconstruction methods for 4x up-sampling on simulated measurements with a high, medium and low signal-to-noise.} RMSE is the root-mean-square error; ADE is the absolute depth error.}
\end{center}
\end {table}



\subsubsection{Robustness to noise}
\begin{figure*}
\includegraphics[width=\textwidth]{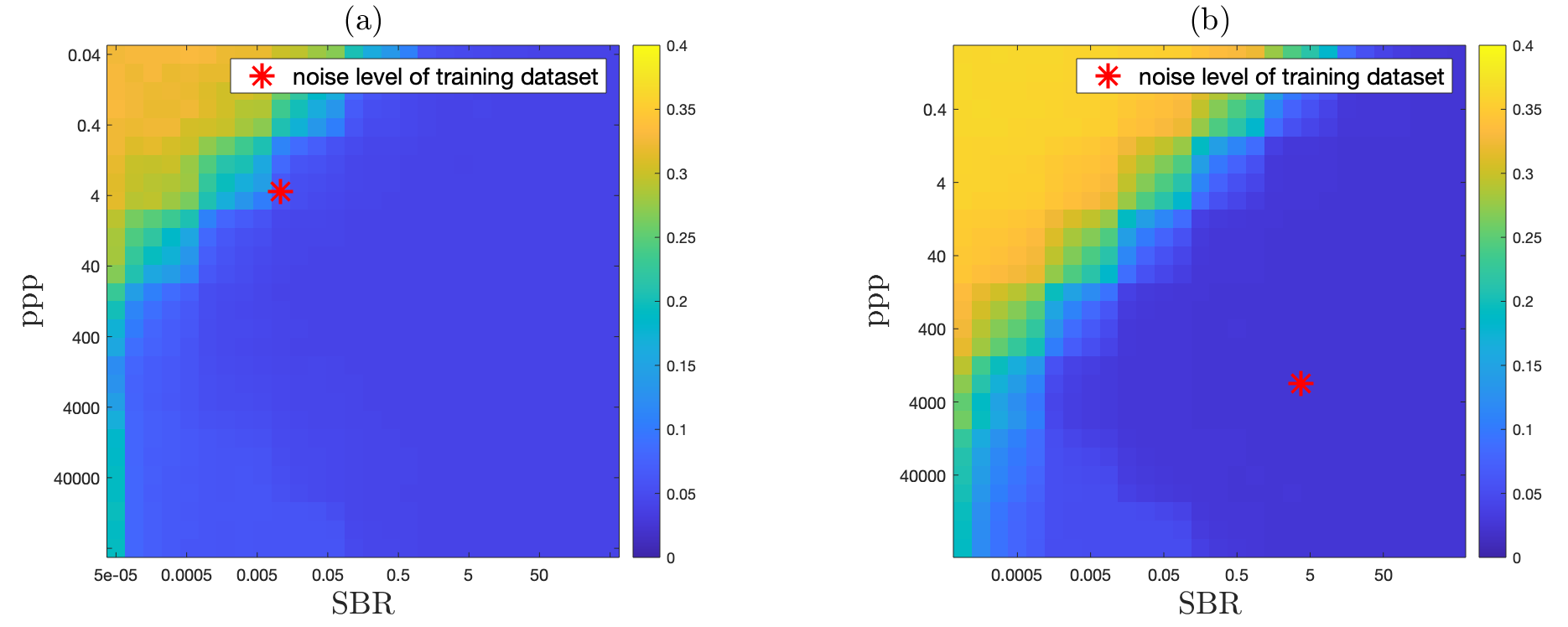}
\caption{\label{fig:robustness} 
\textbf{RMSE between predictions of network and ground truth for different SBR and ppp.} The noise level implemented in the training dataset is represented with a red marker. In (a), medium signal-to-noise, the network was trained on data with an average signal count ($ppp$) of 4 per pixel and a signal to background ratio (SBR) of 0.02. In (b), high-signal-to-noise, the network was trained on data with an average signal count of 1200 and a signal to background ratio (SBR) of 2. Both networks were tested on simulated data with $ppp$ levels ranging from $1*10^{-1}$ to $7*10^{5}$ and SBR ranging from $1*10^5$ to $70$. The network performs best when the training and testing noise matches. It is also robust to data that have a lower noise. However, the performance drops when the testing data presents higher noise than the training data.  
}
\end{figure*}

\begin{figure*}
\begin{center}
\includegraphics[width=\textwidth]{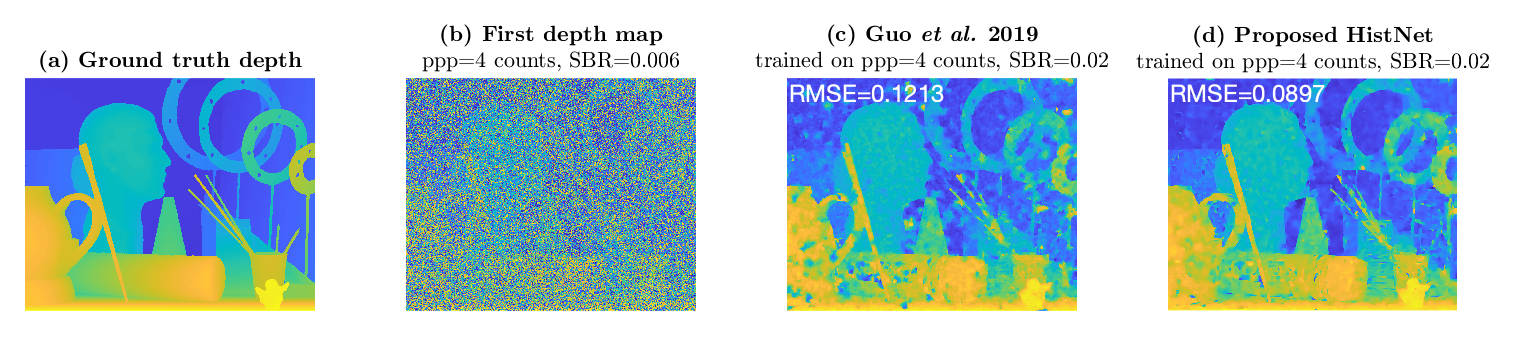}
\caption{\label{fig:robustness_result_fig}\textbf{Reconstruction of data with low signal-to-noise (ppp=4 counts and SBR=0.006) with HistNet and DepthSR-Net of Guo \textit{et al.} \cite{:DepthSR-Net} trained on data with medium signal-to-noise (ppp=4 and SBR=0.02).} (a) is the ground truth depth; (b) shows the first depth map input of the network ; (c) is the 4x-upsampled reconstruction with DepthSR-Net of Guo \textit{et al.} ; (d) s the 4x-upsampled reconstruction with HistNet. RMSE is the root-mean-square error.
}
\end{center}
\end{figure*}

We study how well a network trained on data with a specific SBR and ppp levels can reconstruct data with other noise levels. HistNet trained according to the high and medium signal-to-noise scenarios was tested on data with ppp levels ranging from $1*10^{-1}$ to $7*10^5$ and SBR ranging from $1*10^{-5}$ to $70$. Figure \ref{fig:robustness} shows the RMSE value between the ground truth and the reconstruction of HistNet with respect to the SBR and the $ppp$ of the testing data. 
We see that HistNet shows good performance on a variety of SBR and $ppp$ levels. HistNet trained with the medium signal-to-noise  scenario is able to reconstruct data that presents higher noise than when trained with the high signal-to-noise scenario. However, the best performance is always achieved when the $ppp$ and SBR of the testing data approximately matches the one implemented in the training dataset. Figure \ref{fig:robustness_result_fig} shows the reconstruction of data of low signal-to-noise scenario by HistNet trained on data with median signal-to-noise scenario.


\subsubsection{Ablation study}
In this section, we investigate the relative benefits of  the different input features of the network, i.e.~the intensity image, the multi-scale depth features, and the second depth map. For each of these different feature inputs, we used three versions of HistNet: one that does not use the intensity (w/o Intensity); one that does not use the multi-scale depth features (w/0 Depth Features); and one that does not use the second depth maps (w/0 Second Depth). We trained each of these versions of HistNet for 4x up-sampling for the medium and high noise scenarios. 

Quantitative results of different simulated measurements from the Middlebury dataset are displayed in Tables \ref{tab:tableablation_realistic_noisyscenario} and \ref{tab:tableablation_extreme_noisyscenario} for the high and medium signal-to-noise scenarios, respectively.  In the high signal-to-noise scenario, the absence of the second depth map results in a noticeable reduction in the quality of the reconstruction. In the medium signal-to-noise  scenario, the absence of the multi-scale depth features and the intensity image significantly reduces the performance of the network.  These results demonstrate the value that each of the features brings to the network.

\begin {table}[!ht]
\footnotesize
\begin{center}
\begin{tabular}{|C{1cm}|C{1cm}|C{1cm}|C{1cm}|C{1cm}|C{1.4cm}|C{1cm}|C{1.4cm}|C{1cm}|} 
\hline
 & \multicolumn{2}{c|}{\textbf{HistNet}} & \multicolumn{2}{c|}{\textbf{HistNet w/o Intensity}} & \multicolumn{2}{c|}{\textbf{HistNet w/o Depth Features}} & \multicolumn{2}{c|}{\textbf{HistNet w/o Second Depth}} 
 
 \\ \hline
 \textbf{Scene} & \textbf{RMSE} & \textbf{ADE} & \textbf{RMSE} & \textbf{ADE} & \textbf{RMSE} & \textbf{ADE} & \textbf{RMSE} & \textbf{ADE}  \\ \hline
 Art & 0.0230 & \textbf{0.0027} & \textbf{0.0210}  &  \textbf{0.0027} & 0.0230 & 0.0029 & 0.0260 & 0.0041  \\ \hline
 Books & \textbf{0.0057} & 0.0014 & 0.0061 & 0.0014 & 0.0060 & 0.0014 & 0.0082 & 0.0017  \\ \hline
 Dolls & \textbf{0.0050} & \textbf{0.0015} & 0.0055 & 0.0017 & 0.0051 & 0.0016 & 0.0061 & 0.0020  \\ \hline
 Laundry & 0.0128 & \textbf{0.0033} & \textbf{0.0124} &  0.0034 & 0.0128 & 0.0034 & 0.0150 & 0.0043   \\ \hline
 Moebius & \textbf{0.0129} & \textbf{0.005} & 0.0153 & 0.0064 & 0.0140 & 0.0056 & 0.0150 & 0.0063  \\ \hline
 Reindeer & 0.0126 & 0.0019 & 0.0125 & 0.0019 & \textbf{0.0123} & 0.0019 & 0.0140 & 0.0023  \\ \hline
 AVG & \textbf{0.012 $\pm$ 0.0065} & \textbf{0.026 $\pm$ 0.014} & 0.012 $\pm$0.0058 & 0.0029 $\pm$ 0.0019 & 0.012 $\pm$ 0.0065 & 0.0028 $\pm$ 0.0015 & 0.014 $\pm$ 0.0069 & 0.0095 $\pm$ 0.0153  \\ \hline
\end{tabular}
\caption{\label{tab:tableablation_realistic_noisyscenario} \textbf{High signal-to-noise. Quantitative comparison on different simulated measurements with ppp = 1200 counts and SBR = 2 between versions of HistNet with and without using the intensity maps, the down-sampled features or the second depth map.} The version of HistNet that integrates the intensity, the down-sampled features and the second depth map has the best performance in terms of RMSE and ADE for most of the simulated measurements. RMSE is the root-mean-square error; ADE is the absolute depth error. }
\end{center}
\end {table}

\begin {table}[!ht]
\footnotesize
\begin{center}
\begin{tabular}{|c|c|c|c|c|c|c|} 
\hline
  & \multicolumn{2}{c|}{\textbf{HistNet}} & \multicolumn{2}{c|}{\textbf{HistNet w/o Intensity}} & \multicolumn{2}{c|}{\textbf{HistNet w/o Depth Features}} \\ \hline
\textbf{Scene} & \textbf{RMSE} & \textbf{ADE} & \textbf{RMSE} & \textbf{ADE} & \textbf{RMSE} & \textbf{ADE} \\ \hline
Art & \textbf{0.053} & \textbf{0.019} & 0.057 & 0.020 & 0.064 & 0.023  \\ \hline
Books & \textbf{0.017} & \textbf{0.009} & 0.021 & 0.010 & 0.019 & 0.010 \\ \hline
Dolls & \textbf{0.019} & \textbf{0.012} & 0.029 & 0.013 & 0.061 & 0.018 \\ \hline
Laundry & \textbf{0.024} & 0.013 & 0.026 &  \textbf{0.012} & 0.026 & 0.013  \\ \hline
Moebius & \textbf{0.025} & 0.015 & 0.024 & \textbf{0.013} & 0.045 & 0.020  \\ \hline
Reindeer & \textbf{0.040} & \textbf{0.019} & 0.050 & 0.021 & 0.055 & 0.023  \\ \hline
AVG & \textbf{0.030 $\pm$ 0.013} & \textbf{0.015  $\pm$ 0.004} & 0.034 $\pm$ 0.014 & \textbf{0.015  $\pm$ 0.004} & 0.045 $\pm$ 0.017 & 0.018 $\pm$ 0.005  \\ \hline
\end{tabular}
\caption{\label{tab:tableablation_extreme_noisyscenario} \textbf{Medium signal-to-noise. Quantitative comparison on different simulated measurements with average ppp = 4 counts and SBR = 0.02 between versions of HistNet with and without using the intensity maps, the down-sampled features or the second depth maps.} The version of HistNet that integrates both the intensity and the down-sampled features performs better than the one without it in terms of RMSE and ADE for most of the simulated measurements.  RMSE is the root-mean-square error; ADE is the absolute depth error.}
\end{center}
\end {table}

\subsection{Results for the Quantic 4x4 camera (4-times up-sampling)}

\subsubsection{Quantic 4x4 camera}
\begin{figure*}
\includegraphics[width=\textwidth]{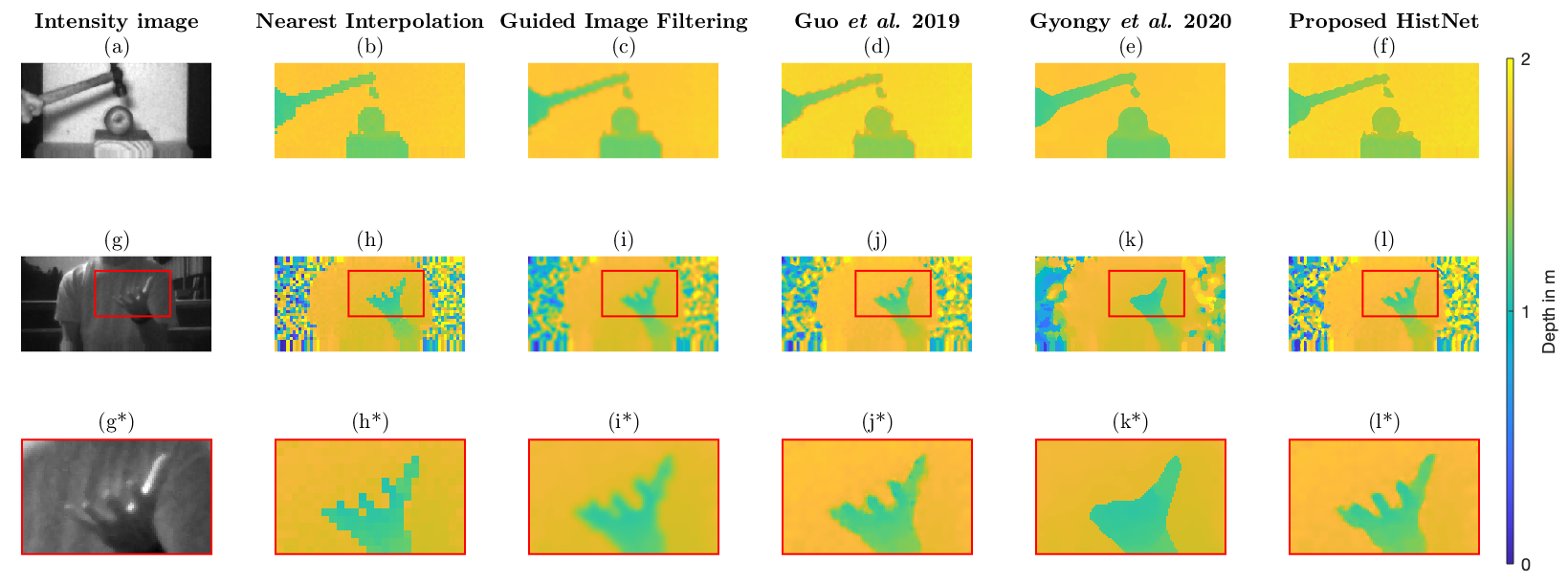}
\caption{\label{fig:Compa_real} \textbf{Reconstruction of Quantic 4x4 data.}
(a) displays the reflectivity image from the SPAD \cite{:GyongyOptica}; (b) is the reconstruction of the depth using nearest interpolation ; (c) is the reconstruction of the depth using guided image filtering \textit{et al.} 2013 \cite{guidedfiltering}; (d) is the reconstruction via the algorithm DepthSR-Net of Chunle Guo \textit{et al.} 2019 \cite{:DepthSR-Net}; (e) displays the reconstruction of the algorithm of Gyongy \textit{et al.} 2020 \cite{:GyongyOptica} ; (f) is the reconstruction of our proposed HistNet. The second row (g)-(l) displays the equivalent for another captured measurement and (g*)-(l*) displays closeup of (g)-(l).}
\end{figure*}

We test the performance of HistNet on real measurements captured by the Quantic 4x4 camera \cite{:GyongyOptica}. The spatial resolution of the histogram data is of 32x64 and the number of time bins is of 16. The resolution of the intensity image is of 128x256. The data is first interpolated to the size of the intensity image with nearest interpolation and is calibrated using a compensation frame. 
We estimated a number of photon counts per pixel of 1200 and a signal-to-background ratio of 2 in this data. Therefore, we use HistNet trained with this scenario to reconstruct the depth maps. Figure \ref{fig:Compa_real} show the reconstruction of Quantic 4x4 data via HistNet together with a comparison with different reconstruction algorithms. We see that HistNet leads to more accurate image with sharper edges. 

\newpage

\subsection{Generalisation to a new dataset (8-times up-sampling)} \label{8times}
In this section, we consider the case of 8-fold up-sampling of depth from data acquired by another single-photon detector array. Here we apply  our algorithm for up-sampling  the dataset collected and  presented in Lindell \textit{et al.} \cite{:Lindell}.  This shows that our network can be applied to other data formats, not just data from the Quantic 4x4 camera.

\subsubsection{Description of the data}
The SPAD array sensor in ref \cite{:Lindell} acquires histograms of photon counts with a spatial resolution of 72x88 with 1024 time bins. 
A conventional high-resolution camera acquires a corresponding HR intensity image at a resolution of 576x704 pixels (8x higher spatial resolution than the original data). The histograms are very sparse, i.e.~the return signal from objects is only contained within a range of 60 time bins, and the remaining 924 time bins contain only background photons.

\subsubsection{Pre-processing of data for the network}

The LR histograms (of size 72x88x1024) are cropped in the temporal dimension to include only the return signal from objects as described in \ref{Outlier_rejection_step}.  We note though that the rejection step removes data in temporal bins where there is no signal, so we do not anticipate that it has any impact on the signal-to-background ratio (\ref{equation_SBR}) where any signal is present.

The first depth map and the multi-scale depth features are computed from this cropped version of the LR histogram. The first depth map is computed using a matched filter on this cropped histogram of photon counts, which is then up-scaled with the nearest neighbour interpolation to the desired 576x704 resolution (8x larger in both spatial dimensions in this section). The first depth map is divided by the number of bins of the cropped histogram to be normalize between zero and one. Unlike the Quantic 4x4, the sensor considered in this section does not integrate the histograms of 4x4 pixels. Therefore, the second depth map is set to zero. 

The  multi-scale depth features D1 and D2 are obtained by down-sampling by two and four the previously obtained first depth map. D3 is obtained by using a matched filter on the cropped histogram of photon counts. D4 is obtained by down-sampling the cropped histogram of photon counts by two by summing the neighbouring pixels in the spatial dimension and by using matched filter on the down-sampled histogram. All D1-4 are divided by the number of bins of the cropped histogram. The dimensions of the features are 288x352, 144x276, 72x88 and 36x44 for D1, D2, D3 and D4 respectively. The HR intensity acquired by the camera is normalized between zero and one.  Note that the algorithm of Lindell {\it et al.}~does not require any pre-processing  as it operates directly on histogram data.

\subsubsection{Training of HistNet for 8-times up-sampling}

We simulate the SPAD measurements (LR Histogram and HR intensity) from  23 depth images extracted from the MPI dataset. The HR depth and intensity images are first down-sampled by a factor of eight with bicubic interpolation. Histograms with 60 time bins are computed from these LR depth and LR intensity images (as detailed in \ref{Training-Dataset}), so that the depth is spread over the entire range of bins. The noise level of our training dataset is chosen to match the noise level of the Lindell dataset as close as possible.  We use a ppp of 2 counts and a SBR of 40.


\subsubsection{Comparison algorithms}

\begin{itemize}
    \item \textit{Nearest-neighbour interpolation}  Depth is estimated with matched filter on the cropped 276x344x60 histogram of counts and up-sampled with nearest neighbour interpolation to a 576x704 image.
    
    \item \textit{Guided Image Filtering of He \textit{et al.}~2013} \cite{guidedfiltering} We perform further processing to the estimated depth from the nearest-neighbour interpolation by applying the guided filtering algorithm with the HR intensity image as a guide.
    
    \item \textit{DepthSR-Net of Guo \textit{et al.}~2019} \cite{:DepthSR-Net} We trained this network using the same training datasets for our network. This network outputs an 8x up-sampled depth map from a LR depth map using a HR intensity map to guide the reconstruction. 

    \item \textit{Network of Lindell et al.~2018} \cite{:Lindell} 
    We use the end-to-end trained network for depth estimation and guided 8x up-sampling presented in \cite{:Lindell}. This network takes a LR histogram of photon counts and a HR intensity image as the input, and it outputs an eight-fold up-sampled depth. The network consists of two parts: first, a denoising branch that estimates a clear depth image from the histogram; and second, a guided up-sampling branch that up-samples the depth image to the desired resolution. Both branches consist of a series of 3D convolutions and make use of the intensity image as guidance. The denoising branch processes the histogram of photon counts at multiple scales.

\end{itemize}

\subsubsection{Results}

\begin{figure*}
\includegraphics[width=\textwidth]{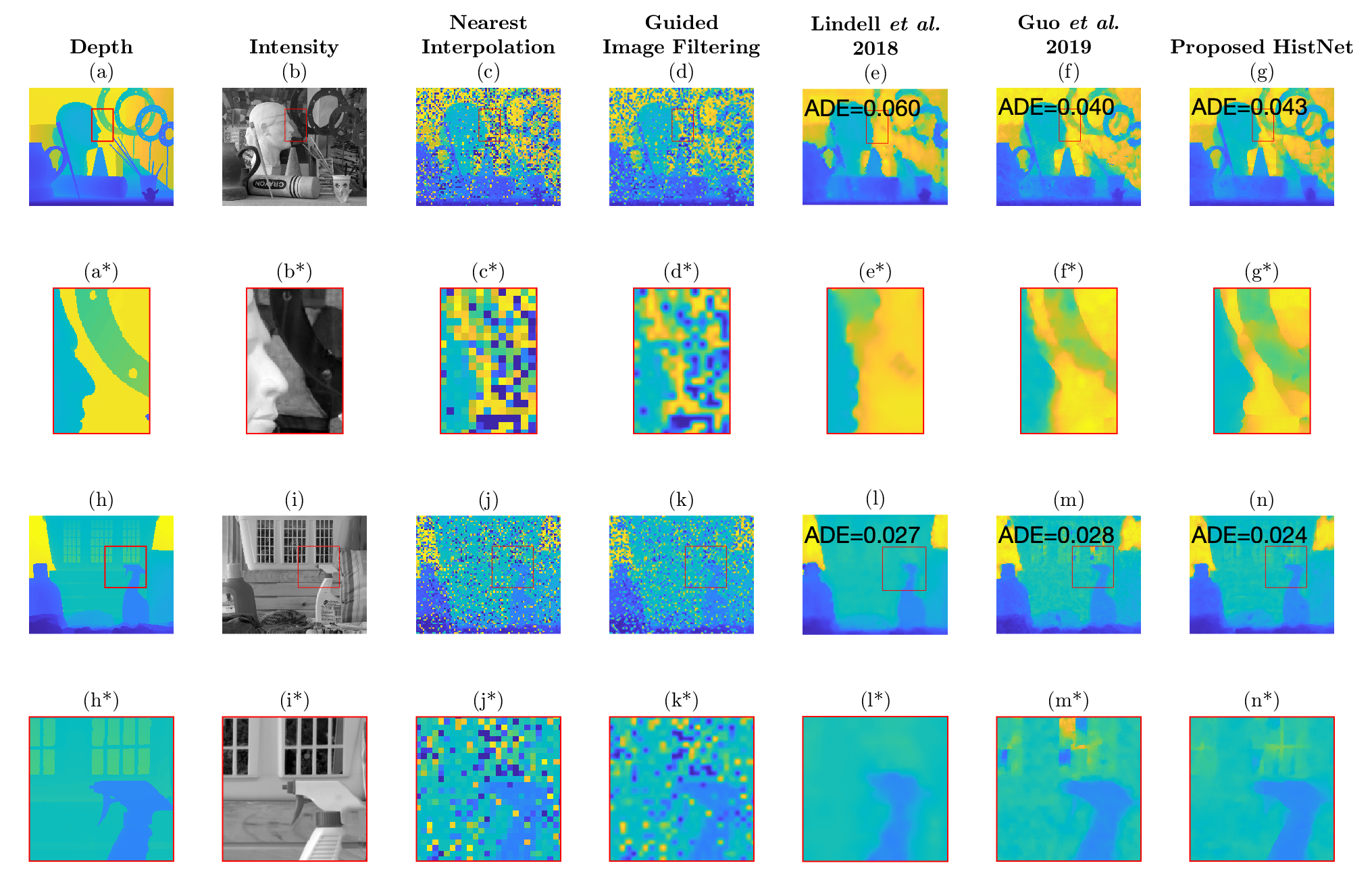}
\caption{\label{fig:Compa_8_fold} \textbf{Comparaison of reconstruction techniques for 8-fold up-sampling}
ADE is the absolute depth error calculated with normalized data between 0 and 1. (a) displays the ground truth depth image; (b) displays the ground truth of the intensity; (c) is the reconstruction of the depth with nearest interpolation ; (d) is the up-sampled version of the first depth data with Guided Image Filtering \cite{guidedfiltering}; (e) is the reconstruction estimated by the CNN network of Lindell \textit{et al.} 2018 \cite{:Lindell}; (f) is the reconstruction of the first depth data with the algorithm DepthSR-Net of Guo \textit{et al.} 2019 \cite{:DepthSR-Net}; (g) is the reconstruction via our proposed method HistNet.  (a*), (b*), (c*), (d*), (e*), (f*) and (g*) are the closeup views of (a)-(g). (h)-(n) are the corresponding images from another simulated measurements with (h*)-(n*) the closeup views.}
\end{figure*}

Figure \ref{fig:Compa_8_fold} shows the reconstruction for the different reconstruction methods, and we report quantitative results for eight Middlebury scenes in Table \ref{tab:table-8x}.  This table shows that HistNet compares well with the state-of-the-art algorithms DepthSR-Net \cite{:DepthSR-Net} and that of Lindell \textit{et al.} \cite{:Lindell}.  The RMSE and ADE metrics are all within one standard deviation of each other, thus all algorithms are extremely close in terms of image quality in this scenario.  This similarity in performance is to be expected due to the high signal-to-noise level of this dataset.   In terms of processing time, HistNet and the network of Guo \textit{et al.} \cite{:DepthSR-Net} take 3.6s to reconstruct the image, the network of  Lindell \textit{et al.} \cite{:Lindell} takes 11.7s and guided filtering and nearest interpolation processes the image in a few milliseconds. The networks were tested using a NVIDIA RTX 6000 GPU. 
We note that our proposed HistNet and DepthSR-Net of Guo \textit{et al.} \cite{:DepthSR-Net} act on pre-processed histograms by extracting informative depth features. Hence, the processing time of those learning-based methods depends on the spatial desired resolution of the depth maps and does not depend on the number of time bins in the histogram.

\begin {table}[!ht]
\label{table:compaLindell}
\footnotesize
\begin{center}
\begin{tabular}{|m{2cm}|m{1cm}|m{1cm}|m{1cm}|m{1cm}|m{1cm}|m{1cm}|m{1cm}|m{1cm}|m{1cm}|m{1cm}|} 
\hline
  & \multicolumn{2}{c|}{\textbf{Proposed HistNet}} & \multicolumn{2}{c|}{\textbf{Lindell \textit{et al.}} \cite{:Lindell}} & \multicolumn{2}{c|}{\textbf{Guo \textit{et al.}} \cite{:DepthSR-Net}} & \multicolumn{2}{c|}{\textbf{NNI}} & \multicolumn{2}{c|}{\textbf{Guided Image Filtering}} \\ \hline
  Rec time per scene & \multicolumn{2}{c|}{3.6s (on GPU)} & \multicolumn{2}{c|}{11.7s (on GPU)} & \multicolumn{2}{c|}{3.6s (on GPU)} & \multicolumn{2}{c|}{0.004s} & \multicolumn{2}{c|}{0.05s} \\ \hline
 \textbf{Scene} & \textbf{RMSE} & \textbf{ADE} & \textbf{RMSE} & \textbf{ADE} & \textbf{RMSE} & \textbf{ADE} &\textbf{RMSE} & \textbf{ADE} & \textbf{RMSE} & \textbf{ADE}\\ \hline
Art & 0.092 &  0.043 & 0.111 & 0.060 & \textbf{0.085} & \textbf{0.040} & 0.313 & 0.182 & 0.264 & 0.169 \\ \hline
Books &  0.042 &  0.024 & \textbf{0.039} & \textbf{0.022} & 0.042 & 0.023 & 0.315 & 0.177 & 0.265 & 0.165\\ \hline
Dolls & 0.046 &  0.024 &  \textbf{0.037} &  \textbf{0.024} & 0.039 & 0.023 & 0.280 & 0.155 & 0.227 & 0.138 \\ \hline
Laundry & 0.056 &  \textbf{0.024} &  \textbf{0.052} & 0.027 & 0.062 & 0.028 & 0.232 & 0.128 & 0.185 & 0.112 \\ \hline
Moebius & 0.042 & \textbf{0.024} &  0.043 &  0.027 & \textbf{0.039} & \textbf{0.024} & 0.273 & 0.152 & 0.221 & 0.139 \\ \hline
Reindeer &  0.110 & 0.067 & \textbf{0.085} & \textbf{0.043} & 0.117 & 0.067 & 0.383 & 0.230 & 0.334 & 0.213 \\ \hline
Bowling &  \textbf{0.065} &  \textbf{0.028} & 0.070 & 0.030 & 0.074 & 0.035 & 0.274 & 0.157 & 0.222 & 0.139\\ \hline
Plastic & 0.046 & 0.023 &  \textbf{0.037} &  \textbf{0.019} & 0.052 & 0.026 & 0.252 & 0.140 & 0.197 & 0.121 \\ \hline
AVG & 0.062 $\pm$ 0.023 & 0.032 $\pm$ 0.015 & \textbf{0.059 $\pm$ 0.027} & \textbf{0.031 $\pm$ 0.013} & 0.064 $\pm$ 0.026 & 0.033 $\pm$ 0.014 & 0.290 $\pm$ 0.047 & 0.165 $\pm$ 0.032 & 0.240 $\pm$ 0.047 & 0.150 $\pm$ 0.032 \\ \hline
\end{tabular}
\caption{\label{tab:table-8x} \textbf{Quantitative comparison of the different reconstruction methods for 8x up-sampling on simulated measurements with a ppp of 2 counts and an SBR of 40.} RMSE is the root-mean-square error; ADE is the absolute depth error. The error metrics were calculated on normalized data such that range of the ground truth depth images was between zero and one. }
\end{center}
\end {table}

\newpage
\section{Discussion and conclusion}
\label{conclusion}
In this work we present a deep network for up-scaling and denoising depth images. The network is designed for measurements provided by SPAD array sensors and provides significant improvements for image quality and resolution over a wide range of noise scenarios. Our method exploits the SPAD array data in a simple and efficient manner, i.e.~we extract multi-scale features and multiple depths from the histogram data, and these are provided directly to the network.  Additionally we made use of the $l1$ loss to promote sparsity in the reconstructed images. 

The combination of the additional features and the loss function result in a network with state-of-the-art performance in terms of image metrics and processing times. The network performs well with respect to other up-sampling algorithms, especially when applied to data with low signal-to-noise ratios and low photon levels. Moreover, our method shows robustness to a wide range of different noise scenarios, so that the noise statistics of the training dataset do not need to closely match the input data's.   Future work will focus on the high frame rate of the SPAD array sensor and use information in the temporal domain to achieve better spatial resolutions for depth images.  We also propose to tackle the misalignment between the histogram and the intensity image, which is inherent to the operating mode of our SPAD detector that acquires them alternately. 

\section*{Funding}

Engineering and Physical Science Research Council (EP/T00097X/1, EP/S001638/1 and EP/L016753/1);
UK Royal Academy of Engineering through the Research Fellowship Scheme under Grant RF/201718/17128;
DSTL Dasa project DSTLX1000147844

\section*{Acknowledgements}

The authors are grateful to STMicroelectronics and the ENIAC-POLIS project for chip fabrication for the Quantic 4x4 sensor.  We are grateful to the authors of the DepthSRNet paper for sharing their code \cite{:DepthSR-Net}.  The code for this work can be found at https://github.com/HWQuantum/HistNet. We thank Fr\'ed\'eric Ruget for helpful discussions relating to the simulation of the datasets.

\section*{Disclosures}

The authors declare no conflicts of interest.

\end{document}